\documentclass{article}

\usepackage{PRIMEarxiv}
\usepackage[utf8]{inputenc} 
\usepackage[T1]{fontenc}    
\usepackage{hyperref}       
\usepackage{url}            
\usepackage{booktabs}       
\usepackage{amsfonts}       
\usepackage{nicefrac}       
\usepackage{microtype}      
\usepackage{lipsum}
\usepackage{fancyhdr}       
\usepackage{graphicx}       
\graphicspath{{media/}}     
\usepackage{comment}
\title{Structure Preserving Cycle-GAN for Unsupervised Medical Image Domain Adaptation
}

\author{
  Paolo Iacono, Naimul Khan \\
  Department of Electrical and Computer Engineering \\
  Toronto Metropolitan University \\
  Toronto\\
  \texttt{\{piacono, n77khan\}@torontomu.ca} \\
}

\begin{document}
\maketitle

\begin{abstract}
The presence of domain shift in medical imaging is a common issue, which can greatly impact the performance of segmentation models when dealing with unseen image domains. Adversarial-based deep learning models, such as Cycle-GAN, have become a common model for approaching unsupervised domain adaptation of medical images. These models however, have no ability to enforce the preservation of structures of interest when translating medical scans, which can lead to potentially poor results for unsupervised domain adaptation (DA) within the context of segmentation. This work introduces the Structure Preserving Cycle-GAN (SP Cycle-GAN), which promotes medical structure preservation during image translation through the enforcement of a segmentation loss term in the overall Cycle-GAN training process. We demonstrate the structure preserving capability of the SP Cycle-GAN both visually and through comparison of Dice score segmentation performance for the unsupervised domain adaptation models. The SP Cycle-GAN is able to outperform baseline approaches and standard Cycle-GAN domain adaptation for binary blood vessel segmentation in the STARE and DRIVE datasets, and multi-class Left Ventricle and Myocardium segmentation in the multi-modal MM-WHS dataset. SP Cycle-GAN achieved a state of the art Myocardium segmentation Dice score (DSC) of 0.7435 for the MR to CT MM-WHS domain adaptation problem, and excelled in nearly all categories for the MM-WHS dataset.  SP Cycle-GAN also demonstrated a strong ability to preserve blood vessel structure in the DRIVE to STARE domain adaptation problem, achieving a 4\% DSC increase over a default Cycle-GAN implementation. 

\end{abstract}

\keywords{Domain Adaptation \and Unsupervised \and MRI \and CT \and Retinal Scan \and STARE \and DRIVE \and Segmentation \and MM-WHS}

\section{Introduction}

Domain shift is a common and difficult problem to deal with in medical image processing. Medical datasets can have a large variance in image domain similarity due to difference in imaging techniques, different scanner equipment, etc. Medical image analysis, including image classification and segmentation, has been shown to suffer degraded results in cases where the datasets suffer from a domain shift in the images used \cite{Guan_2022}. For example, Brain MRI scans from different scanners can have completely different intensity distributions, which can hurt models using these images in a combined setting, as the models are not considering the presence of this domain shift during training \cite{Guan_2022}. The field of domain adaptation (DA) involves the development of algorithms or models which can correct for this domain shift in data. 

Addressing this problem with a pixel-level machine learning approach, such as the use of Cycle-GANs for image domain adaptation, has been shown to be an effective solution for domain shift in medical images \cite{MedCGAN}. Use of Cycle-GAN for domain adaptation of brain lesion MR scans has been shown to reduce the Jenson-Shannon divergence between image domains, which allows for predictive models to operate on unlabelled domains which the model has not previously encountered \cite{MedCGAN}. However, the use of a traditional Cycle-GAN for unsupervised domain adaptation in medical applications may not sufficiently preserve important structures in the images, which is obviously important for medical segmentation tasks. 

In this work, we devised a simple segmentation-based loss function for Cycle-GAN training, using simultaneous U-Net model training during Cycle-GAN training in an effort to enforce the preservation of the structures of interest during image translation. This Structure Preserving Cycle-GAN (SP Cycle-GAN) can preserve the structure of organs or areas in scans which are segmented after the scan is translated to another domain. This structure preservation in turn improves segmentation performance over the standard Cycle-GAN methods of domain adaptation for medical segmentation datasets, including Left Ventricle (LV) and Myocardium (Myo) segmentation from CT/MR scans from the MM-WHS dataset \cite{CFDistance}, and blood vessel segmentation from retinal scans for the STARE and DRIVE datasets \cite{STARE,DRIVE}. 

\subsection*{Generalizable Insights about Machine Learning in the Context of Healthcare}

This work demonstrates that the simple inclusion of a segmentation loss into adversarial-based methods for unsupervised medical image domain adaptation can improve segmentation results significantly for the translated images. This could potentially be extended to other methods of deep domain adaptation apart from the Cycle-GAN based method featured in this experiment, and provide more robust unsupervised domain adaptation models for clinicians to use in the context of medical image segmentation. The domain adaptation approach highlighted in this work can be used when training data from a target domain is difficult to obtain due to resource limitations, such as remote or marginalized communities.

\section{Literature Review}
As mentioned earlier, often it is common for different portions of data to suffer from domain shift, where data is split into multiple different domains with distinct distributions and feature domains \cite{DA_Surv}. This domain shift issue can lead predictive models trained on one domain to struggle when it is exposed to unseen data in a different domain. The main concept of domain adaptation is to learn a model which can generalize labelled source domain data to a target domain by minimizing the distribution differences between the domains \cite{DA_Surv}. Several approaches to domain adaptation have been devised in the past, depending on the form of the data. The problem of domain shift can be categorized into three sections: prior shift (class imbalance), covariate shift, and concept shift \cite{Moreno}. Most of the attempts to create domain adaptation models focus on the covariate shift problem, which is defined when two probability distributions differ, while the conditional probability distributions are equivalent across domains \cite{DA_Surv}. 

Domain adaptation falls into several approach categories, including instance-based adaptation, feature-based adaptation and deep domain adaptation \cite{DA_Surv}. The focus of this work is on deep domain adaptation, which revolves around the use of neural deep learning models for domain adaptation, and more specifically on the adversarial approach to deep domain adaptation. Deep adversarial domain adaptation is mainly inspired by the use of Generative Adversarial Networks (GANs), which are deep-learning models which learn based on a two player game, where a Generator creates an artificial image in the target domain which attempts to fool the Discriminator into misclassifying it as a sample of the ground truth target domain class \cite{GAN}. In effect, GANs in a domain adaptation application are attempting to convert images in the source domain to images with as small a discrepancy as possible to the target domain \cite{DA_Surv}, through a pixel-level image translation process. Attempts to build GANs for domain adaptation include Domain Transfer Network \cite{Domain_Transfer} and Cycle-GAN \cite{Cycle}, the latter of which inspired the domain adaptation approach for this work. 

Domain adaptation has an important role in the medical image analysis, due to the common presence of domain shift in medical data and images \cite{Guan_2022}. This domain shift is caused by factors such as the use of different imaging technologies, the variable properties of different scanners, use of different scanner protocols or even different subject groups \cite{Guan_2022}. These factors can often result in a large domain shift in medical images, for example, brain MR scans using T1-weighted scans vs. T2-weighted scans can cause a large intensity distribution difference in scans \cite{Guan_2022}. Additionally, different imaging modalities such as contrast enhancing T1 (ceT1) or high-resolution T2 (hrT2) images can result in a large domain shift between images \cite{Dorent_2023}. The categorization of the domain adaptation approach taken is based on factors such as label availability, presence of cross-modality, and model type. For example, domain adaptation can be supervised, semi-supervised and unsupervised depending on label availability and the use of ground truth data during the domain adaptation model training \cite{Guan_2022}. Domain adaptation is important in the field of medical image analysis because the domain shift problem can greatly impact predictive model efficacy on unseen image domains \cite{Guan_2022}. An example of the efficacy of GANs for medical domain adaptation was highlighted in the Cross-Modality Domain Adaptation for Medical Image Segmentation (CrossMoDa) 2021 challenge, where GAN-style domain adaptation approaches were effective in domain adaptation between ceT1 and hrT2 style brain MR scans for Vestibular Schwannoma and Choclea segmentation \cite{Dorent_2023}. 

The ability to account for domain shift between medical datasets can allow for predictive models to be trained on imaging modalities that are cheaper, safer and more accessible to clinicians in remote or developing communities. For example, hrT2 brain MRI scans are much more cost effective \cite{Khawaja} as well as less intrusive than ceT1 scans, which require administration of a contrast enhancing agent to the patient \cite{Coelho}. 

Cycle-GAN was introduced in Zhu et al.'s, "Unpaired Image-to-Image Translation using Cycle-Consistent Adversarial Networks" \cite{Cycle}. This paper presented a cyclical GAN model which translates images from a source domain \textit{X} to a target domain \textit{Y}, without the use of a paired image dataset \cite{Cycle}. The proposed method implements two mapping functions, \textit{G}  and \textit{F}, which maps the domain \textit{X} to \textit{Y} and the domain of \textit{Y} to \textit{X}, respectively \cite{Cycle}. The authors used adversarial loss terms for Discriminators \textit{$D_Y$} and \textit{$D_X$} which attempt to classify translated images from \textit{G}  and \textit{F} versus their source domain counterparts. Additionally, the authors posit that learned mapping functions for image translation should by cycle-consistent, such that for a source domain image \textit{x} and domain mapping functions  \textit{F} and \textit{G}, \textit{x} $\rightarrow$ \textit{G(x)} $\rightarrow$ \textit{$F(G(x))$} $\approx$ \textit{x}. This culminates in a model architecture that is capable of bi-directional image translation between two domains of images. The Cycle-GAN loss combining the discussed adversarial and cyclical losses is defined as \cite{Cycle}:

\begin{equation}
\label{eq:CycleGAN}
    L(G,F,D_X,D_Y) = L_{GAN}(G,D_Y,X,Y) + L_{GAN}(F,D_X,X,Y) + L_{cyc}(G,F)
\end{equation}

Where the $L_{GAN}$ terms in Equation \ref{eq:CycleGAN} denote the Discriminator-based Mean Squared Error loss for the X and Y domains, and the $L_{cyc}$ term is an L1 norm loss based on the difference between original and recovered source and target domain images to enforce cyclical consistency \cite{Cycle}. 
Cycle-GAN based DA techniques have been shown to be effective approaches on medical segmentation problems \cite{Dorent_2023}, including DA between STARE and DRIVE datasets for unsupervised DA for medical segmentation \cite{CANNY-GAN}. 

 Vo and Khan (2021) introduced an edge-preserving loss function for Cycle-GAN based medical image domain adaptation for STARE and DRIVE segmentation. The work introduced the EdgeCycleGAN model which preserved the original Cycle-GAN architecture, but added an edge preserving term to the loss function of Equation \ref{eq:CycleGAN}, by introducing an L1 norm of the difference between the canny edge extractor (\textit{$C_e$}) output of the recovered and original input image for each domain of the Cycle-GAN training process \cite{CANNY-GAN}. This resulted in increased segmentation performance on unsupervised DA for blood vessel segmentation over the baseline method without any domain adaptation, and increased performance over the use of a standard Cycle-GAN \cite{CANNY-GAN}. 

In this work we introduce SP-Cycle GAN for structure preserving medical image DA, using simultaneous U-Net segmentation training during Cycle-GAN training to enforce a segmentation loss term in an effort to preserve structures of interest during image translation. This approach does not use any ground truth labels from the target domain, thus making the SP Cycle-GAN an unsupervised DA approach for preserving medical structures during image translation.

\section{Methods}
\subsection{SP Cycle-GAN}
The Structure Preserving Cycle-GAN (SP Cycle-GAN) was implemented by adding a structure preserving loss to the original Cycle-GAN loss shown in Equation \ref{eq:CycleGAN} \cite{Cycle}. This process was inspired by Shin et al. (2021), who implemented a modified Cycle-GAN architecture for unsupervised ceT1 to hrT2 brain MR scan domain adaptation, which preserved Cochlea and Vestibular Schwannoma structures by performing segmentation of the source domain images during training using the accompanying segmentation labels \cite{Shin_Crossmoda}. The SP Cycle-GAN of this experiment did not modify any component of the original Cycle-GAN architecture. Instead, it simply appended a structure preserving term, based on the Focal-Tversky loss between the predicted segmentation result from the recovered source domain image and the original ground truth segmentation label of the source domain image. Focal-Tversky loss is a segmentation loss function introduced by Abraham and Khan (2018), which combines the Tversky index loss function \cite{Tversky} (which is a modification of the Dice loss with adaptation penalization of false negative or false positive pixel classifications) with an exponential focal term, $\gamma$, which focuses the training on lower confidence difficult samples \cite{FTL}. The Tversky index and accompanying Focal-Tversky loss are defined as: 

\begin{equation}\label{eq:TverskyIndex}
    TI_c = \frac{\sum_{i=1}^{N} p_{ic}g_{ic} + \epsilon}{\sum_{i=1}^{N} p_{ic}g_{ic} + \alpha \sum_{i=1}^{N} p_{i\bar{c}}g_{ic} + \beta \sum_{i=1}^{N} p_{ic}g_{i\bar{c}} + \epsilon}
\end{equation}

\begin{equation}\label{eq:FTL}
    FTL_c = \sum_{c}(1-TI_c)^{\frac{1}{\gamma}}
\end{equation}

The $TI_c$ term formulated in Equation \ref{eq:TverskyIndex} is the Tversky similarity index value \cite{Tversky}, which is a weighted form of the Dice Score metric, which allows for flexible penalization of false negative or false positive segmentation classifications. In Equation \ref{eq:TverskyIndex}, $c$ denotes the non-background class to be segmented and $\bar{c}$ denotes the background class, $p_{ic}$ denotes the probability that a predicted pixel $i$ is of the foreground class $c$ and $p_{i\bar{c}}$ is the probability that a predicted pixel $i$ is of the background class. Additionally, $g_{ic}$ denotes the probability that a ground truth pixel $i$ is of the foreground class $c$ and $g_{i\bar{c}}$ is the probability that a ground truth pixel $i$ is of the background class \cite{FTL}. The parameters $\alpha$ and $\beta$ are used for flexible penalization of false negative or false positive classifications \cite{Tversky}. 

As can be seen in Equation \ref{eq:FTL}, the Focal-Tversky loss for segmentation is simply formulated as 1 minus the Tversky similarity index value of Equation \ref{eq:TverskyIndex}, raised to the power of $\frac{1}{\gamma}$. This $\frac{1}{\gamma}$ term is known as the Focal term, and the goal of this term is to reshape the loss function to focus training on low predicted confidence examples which constitute difficult training examples over easily classified training examples \cite{FTL}. 

This Focal-Tversky loss term was added to the original Cycle-GAN loss to create the Structure Preserving Cycle-GAN loss. 

\begin{equation}\label{eq:StructLoss}
    L_{Struct}(U,G,F,X,L) = FTL_c(L,U(F(G(X))))
\end{equation}
\begin{equation}\label{eq:SEGGANLoss}
    L_{SGAN}(G,F,D_X,D_Y) = L_{GAN}(G,D_Y,X,Y) + L_{GAN}(F,D_X,X,Y) \\
    + L_{cyc}(G,F) + \zeta * L_{Struct}(U,G,F,X,L)
\end{equation}
The $L_{Struct}(U,G,F,X,L)$ term shown in Equation \ref{eq:StructLoss} is the structure preserving loss term designed to preserve medical structures during image translation from the SP Cycle-GAN. This function accepts the following parameters: $U$, which is the U-Net segmentation model co-trained during the Cycle-GAN training process that outputs a binary or multi-class segmentation map from an input image, $G$ and $F$, which are Cycle-GAN generators for the source and target domains respectively, $X$, which is an input source domain image, and $L$, which is the accompanying ground truth source domain segmentation label. The $L_{Struct}(U,G,F,X,L)$ loss is calculated as the Focal-Tversky Loss between the predicted  segmentation of the recovered source domain image from the SP Cycle-GAN, $U(F(G(X)))$, and the original ground truth label $X$. The Focal-Tversky loss utilized $\alpha = 0.7$, $\beta = 0.3$ and $\gamma = \frac{4}{3}$. The values of $\alpha = 0.7$, $\beta = 0.3$ were chosen to enforce false negative penalization during training, in an effort to account for the difficult to segment small regions, such as blood vessels. The Focal term $\gamma = \frac{4}{3}$ was selected based on the good segmentation performance this value of $\gamma$ achieved in other experiments using Focal-Tversky loss for lesion segmentation \cite{FTL}. This $L_{Struct}(U,G,F,X,L)$ term is used to train the U-Net segmentation model $U$ during the overall Cycle-GAN training process as well. The overall SP Cycle-GAN loss is shown in Equation \ref{eq:SEGGANLoss}, with $\zeta$ set variably during the experiment based on trial and error testing for each dataset. This $\zeta$ acts as an additional parameter which weighs the influence of the structure preservation via the segmentation loss on the overall GAN training process. 

\begin{figure}
  \centering
  \includegraphics[width=0.95\linewidth]{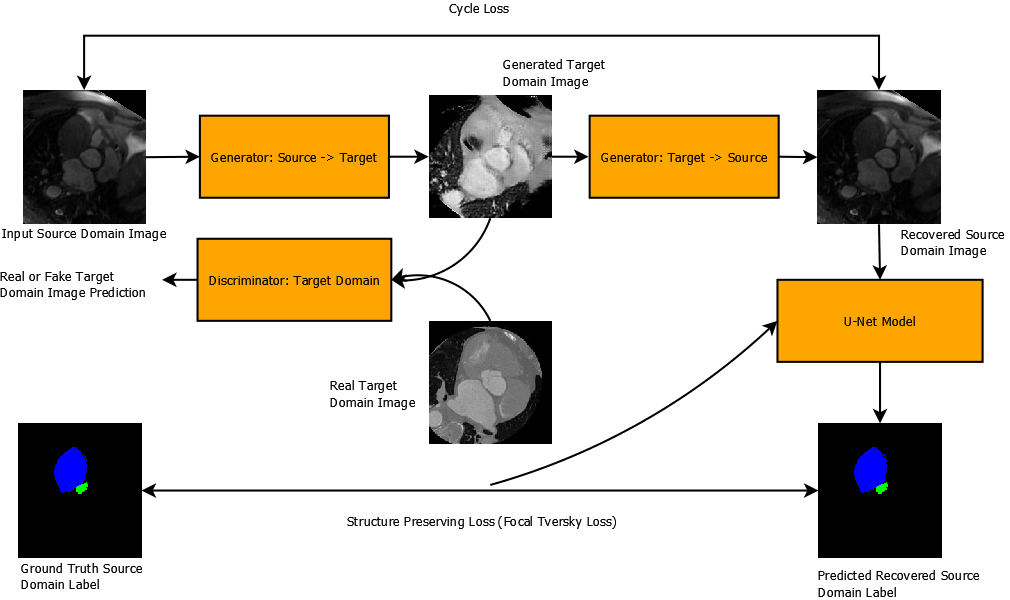}
  \caption{Visualization of Structure Preserving Cycle-GAN training process, using the MM-WHS whole heart segmentation dataset as an example. Where MR is the source domain, and CT is the target domain.}
  \label{fig:SEGGAN}
\end{figure}

The inclusion of this loss was thought to enforce the trained SP Cycle-GAN to preserve and emphasize the important structures in the medical images throughout the cyclical image translation process. The process of the Structure Preserving Cycle-GAN training is shown in Figure \ref{fig:SEGGAN}. 

\subsection{STARE and DRIVE Domain Adaptation}
\begin{figure}
  \centering
  \includegraphics[width=\linewidth]{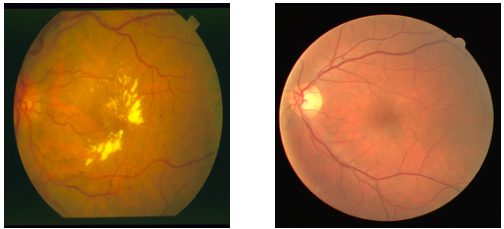}
  
  \caption{STARE (left) vs. DRIVE (right) retina images.}
  \label{fig:STAREvsDRIVE}
\end{figure}

The first domain adaptation datasets used to test the SP Cycle-GAN model are the STARE and DRIVE datasets for blood vessel segmentation \cite{DRIVE,STARE}. The STARE and DRIVE datasets pose an example of the domain shift problem within the area of medical image segmentation. As seen in Figure \ref{fig:STAREvsDRIVE}, the domain shift between these datasets is present in terms of the colour distribution across the scans, as well as the overall shape and structure of the scans. The goal of this experiment was to test the efficacy of the proposed SP Cycle-GAN for preserving the fine structures of the blood vessels in a domain adaptation setting for segmentation for STARE to DRIVE and DRIVE to STARE image translation.  

The STARE dataset features 397 retinal scans, with 20 labelled examples featuring a binary segmentation map denoting blood vessel locations for the respective images \cite{STARE}. The DRIVE dataset features a training set of 20 labelled blood vessel scans and a testing set of 20 labelled blood vessel scans, with 2 sets of manually generated test set labels created by two clinicians \cite{DRIVE}. 

The overall approach for the STARE and DRIVE domain adaptation problem was to first train a SP Cycle-GAN model using the methodology outlined in the previous subsection for both STARE to DRIVE and DRIVE to STARE translation. In an effort to avoid distortion due to the differing shapes of the scans, all scans were first resized to a common size of 512x512 and random crops of 384x384 were taken during each training iteration for the SP Cycle-GAN. The U-Net architecture used for the STARE and DRIVE datasets throughout the experiment was an Attention U-Net model for binary segmentation \cite{AttnUNet}. The Attention U-Net architecture improves over the traditional U-Net model for medical image segmentation \cite{UNet} by featuring a novel attention gate, which is used to highlight important regions of the medical images and suppress features from irrelevant areas of the image. Once the SP Cycle-GAN models are trained for each DA direction, the training set source domain image patches are translated to the target domain. An Attention U-Net model is trained on this translated data, along with the original accompanying ground truth labels. Then, the trained Attention U-Net model is tested on the unseen real target domain test data to evaluate the overall efficacy of the DA approach. 

\subsection{MM-WHS CT and MR Domain Adaptation}
\begin{figure}
  \centering
  \includegraphics[width=\linewidth]{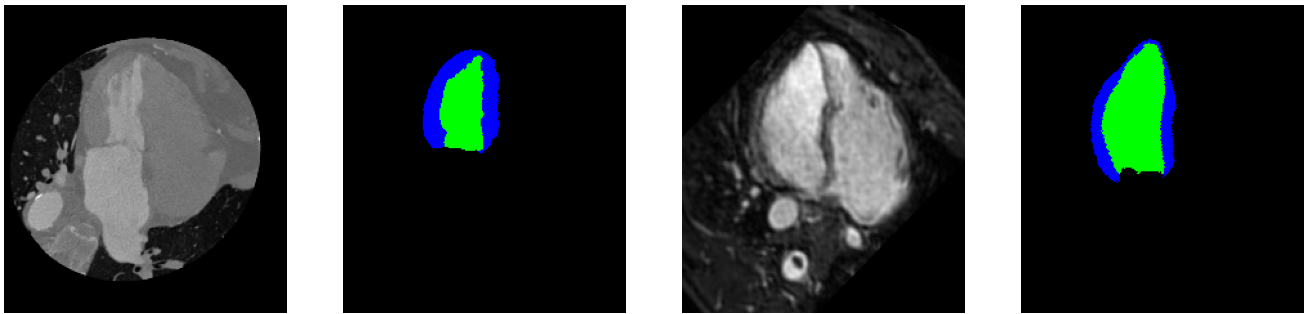}
  
  \caption{CT (first from left) and MR (third from left) 2D slices from the long-axis view centered on the left ventricle, from the MM-WHS Dataset. Accompanying LV and Myocardium labels shown in green and blue respectively to the right of their respective images.}
  \label{fig:MRCT}
\end{figure}

The second DA dataset used in this experiment was a modified version of the MM-WHS dataset \cite{MMWHS} for 3D multi-modal whole heart segmentation, created by Wu and Zhuang (2020). This modified MM-WHS dataset contains 20 cases of labelled CT heart scans, with 16 2D slices from the long-axis view centered on the left ventricle \cite{CFDistance}. It also features 32 unlabelled cases of 16 similar slices for CT. Additionally, this dataset features 20 cases of labelled MR heart scans and 26 unlabelled cases with the same slice configuration as the CT examples. This data features ground truth labels for the Myocardium (Myo) and the Left Ventricle (LV) in both the labelled CT and MR scans, making this a multi-class segmentation problem. An example of the dataset for the MR and CT scans can be seen in Figure \ref{fig:MRCT}. As can be seen in Figure \ref{fig:MRCT}, there is a clear domain shift between the CT and MR heart scans, with a very noticeable difference in pixel intensity distribution and overall structure/shape. 

The SP Cycle-GAN was used to translate images between the MR and CT image domains for unsupervised DA on the modified MM-WHS dataset. In an effort to match the methodology of Wu and Zhuang (2020), images were center-cropped to a fixed size of 192x192 for the SP Cycle-GAN training. Instead of the Attention U-Net model used in the STARE and DRIVE methodology, a U-Net++ model \cite {UNet++} was used for segmentation throughout the MR and CT DA process. 

In a similar fashion to the STARE and DRIVE methodology, a SP Cycle-GAN is trained for the MR to CT and CT to MR DA directions respectively. This training uses the ground truth labelled source domain scans and the unlabelled target domain scans. Then, U-Net++ segmentation models are trained on the translated scans and tested on the unseen ground truth labelled target domain data to evaluate the efficacy of the DA approach. 

\subsection{MM-WHS and STARE/DRIVE SP Cycle-GAN Experiment Setup}
The overall SP Cycle-GAN training process for both the MM-WHS CT/MR and STARE/DRIVE domain adaptation problems follows the workflow shown in Figure \ref{fig:SEGGAN}. The training set-up and details for the STARE/DRIVE and MM-WHS CT/MR SP-Cycle-GAN models are shown in Table \ref{tab:SEGGANTrain}. 

\begin{table}[h!]
    \begin{center}
    \caption{Training Details for SP Cycle-GAN Models for each DA problem.}
    \label{tab:SEGGANTrain}
    \resizebox{\linewidth}{!}{\begin{tabular}{c c c c}

    \hline
    DA Direction & SP Cycle-GAN Image Processing & $L_{Struct}$ Parameters & U-Net Model\\
    
    \hline
    $STARE \rightarrow DRIVE$ & Resize: 512x512, Random Crop: 364x364 & $\alpha = 0.7$ $\beta = 0.3$ $\gamma = \frac{4}{3}$, $\zeta = 3.0$ & Attention U-Net\\
    $DRIVE \rightarrow STARE$ & Resize: 512x512, Random Crop: 364x364 & $\alpha = 0.7$ $\beta = 0.3$ $\gamma = \frac{4}{3}$, $\zeta = 4.0$ & Attention U-Net\\
    $CT \rightarrow MR$ (MM-WHS) & Center Crop: 192x192 & $\alpha = 0.7$ $\beta = 0.3$ $\gamma = \frac{4}{3}$, $\zeta = 5.0$ & U-Net++\\
    $MR \rightarrow CT$ (MM-WHS) & Center Crop: 192x192 & $\alpha = 0.7$ $\beta = 0.3$ $\gamma = \frac{4}{3}$, $\zeta = 4.0$ & U-Net++ \\

    \hline
       
    \end{tabular}}
    \end{center}
    \end{table}

The $\zeta$ parameter value for each DA direction was determined through trial and error, on which value gave the best  results in the segmentation test on unseen testing data. Also, for each DA direction, a default Cycle-GAN with no structure-preserving loss term is trained for comparison with the same training hyper-parameters, image pre-processing and training data. For all SP Cycle-GANs and standard Cycle-GANs trained, a learning rate of 0.0002 is used. For STARE/DRIVE training, the models are trained for a total of 200 epochs with linear annealing to an learning rate of 0 occurring from epoch 150, with a batch size of 2. For MM-WHS training, the models are trained for a total of 100 epochs with linear annealing to an learning rate of 0 occurring from epoch 50, with a batch size of 8. For STARE/DRIVE DA, the DRIVE training set and STARE labelled set, which both feature 20 images and labels, are used for Cycle-GAN training, with the target domain not using any ground truth labels. For MM-WHS DA, the MR domain has 320 labelled and 416 unlabelled images, and the CT domain has 320 labelled and 512 unlabelled images. In both the MR to CT and CT to MR DA directions, the source domain uses  the ground truth labelled set of data and the target domain uses the unlabelled data. A learning rate of 0.001 is used for all U-Net models in this experiment, with the SP Cycle-GAN also linearly annealing the U-Net learning rate at the same time as the Cycle-GAN rate. 

Once the Cycle-GAN models are trained for image translation, the labelled source domain data for each DA problem is translated to the target domain using the SP Cycle-GANs. For STARE/DRIVE, each labelled image is cropped into 4 patches of 364x364 and translated, while the MM-WHS images are centre cropped to 192x192 and translated. The translated images are used to train a U-Net segmentation model, matching the model used by each approach in Table \ref{tab:SEGGANTrain}. To serve as a baseline for comparison to the SP Cycle-GAN and default Cycle-GAN DA approaches, a segmentation model is trained using untranslated data from the source domain as well for each DA direction. The trained segmentation models are tested on the unseen labelled real target domain images and labels, to evaluate the efficacy of the unsupervised DA approaches tested in this work. Table \ref{tab:UNetTrain} specifies the details of the training and testing data for the segmentation models for each DA problem in this work. 

\begin{table}[h!]
    \begin{center}
    \caption{Details for segmentation models trained on translated images from SP Cycle-GAN for DA performance evaluation.}
    \label{tab:UNetTrain}
    \resizebox{\linewidth}{!}{\begin{tabular}{c c c c}

    \hline
    DA Direction & Training Set & Testing Set & U-Net Model\\
    
    \hline
    $STARE \rightarrow DRIVE$ & 80 Translated $STARE \rightarrow DRIVE$ Patches & 80 Test Set DRIVE Patches & Attention U-Net\\
    $DRIVE \rightarrow STARE$ & 80 Translated $DRIVE \rightarrow STARE$ Patches & 80 Test Set STARE Patches & Attention U-Net\\
    $CT \rightarrow MR$ (MM-WHS) &  320 Translated $CT \rightarrow MR$ Images & 320 Test Set MR Images & U-Net++\\
    $CT \rightarrow MR$ (MM-WHS) &  320 Translated $MR \rightarrow CT$ Images & 320 Test Set CT Images & U-Net++\\

    \hline
       
    \end{tabular}}
    \end{center}
    \end{table}

\section{Results}
After the training of the SP Cycle-GAN models defined in Table \ref{tab:SEGGANTrain}, along with the default Cycle-GAN models for the same DA directions for comparison, the trained models are used to translate images for DA evaluation. Segmentation models according to Table \ref{tab:UNetTrain} are trained for images translated using SP Cycle-GAN DA, default Cycle-GAN DA, and no DA, which leads to 3 final trained models to evaluate segmentation performance for each DA direction.
\subsection{Evaluation Approach} 
As mentioned earlier, two datasets are used for DA for segmentation in this experiment, the STARE/DRIVE retinal scan datasets for binary blood vessel segmentation and MM-WHS CT/MR dataset for multi-class LV and Myocardium segmentation. The four unique DA approaches between these sets of data (STARE to DRIVE, DRIVE to STARE, CT to MR and MR to CT), produces 3 final trained segmentation models for each DA direction. This includes the SP Cycle-GAN DA method, default Cycle-GAN DA method, and an untranslated source domain trained segmentation model which serves as a baseline to determine overall DA efficacy for the data. For the all tested DA problems, the overall efficacy of the DA pipeline is determined by the  mean Dice Score on the testing data detailed in Table \ref{tab:UNetTrain}. In the multi-class case of MM-WHS LV and Myocardium segmentation, the mean Dice Score is recorded for both of the non-background classes, as well as an overall average score for all classes. 

\begin{equation}\label{eq:DICE_Eq}
    D(y,\tilde{y}) = 2\frac{|y \cap \tilde{y}|}{|y|+|\tilde{y}|}
\end{equation}

The Dice Score (DCS) formulated in Equation \ref{eq:DICE_Eq} takes ground truth and predicted segmentation maps, $y$ and $\tilde{y}$, as inputs; which are binary pixel arrays predicting the foreground class \cite{DICE}. It is a ratio of the intersection of the two segmentation maps over the addition of the segmentation maps. The DCS outputs a value between [0,1] for the input predicted and ground truth binary segmentation maps, with 1.0 being a pixel-perfect segmentation prediction. In the multi-class setting for the MM-WHS dataset, the DSC defined in Equation \ref{eq:DICE_Eq} is simply calculated for each of the 3 classes being predicted for the image (background, LV, Myocardium). 

\subsection{Results on STARE/DRIVE Domain Adaptation for Binary Blood Vessel Segmentation} 

The final segmentation results for the STARE/DRIVE Domain Adaptation approaches in this experiment are tabulated in Table \ref{tab:STAREDRIVEResults}. 

\begin{table}[h!]
    \caption{Final segmentation results for Unsupervised STARE/DRIVE DA approaches. Best results highlighted in bold.}
    \centering
    \begin{tabular}{c c c}

    \hline
    DA Direction & DA Method & Mean DSC \\
    
    \hline
    $STARE \rightarrow DRIVE$ & SP Cycle-GAN & 0.7679 $\pm$ 0.0033 \\
    $STARE \rightarrow DRIVE$ & Default Cycle-GAN & 0.7455 $\pm$ 0.0040 \\ 
    $STARE \rightarrow DRIVE$ & \textbf{No Domain Adaptation} & \textbf{0.7706} $\pm$ 0.0035 \\ 
    \midrule
    $DRIVE \rightarrow STARE$ & \textbf{SP Cycle-GAN} & \textbf{0.7102} $\pm$ 0.0121 \\
    $DRIVE \rightarrow STARE$ & Default Cycle-GAN & 0.6714 $\pm$ 0.0112 \\
    $DRIVE \rightarrow STARE$ & No Domain Adaptation & 0.6615 $\pm$ 0.0223\\
    \hline
       
    \end{tabular}
    \label{tab:STAREDRIVEResults}
    
\end{table}

As can be seen, for the STARE to DRIVE DA approaches, neither the SP Cycle-GAN nor the default Cycle-GAN DA approaches could outperform the baseline method of simply training on untranslated STARE data.  For the DRIVE to STARE DA approaches, the SP Cycle-GAN method significantly outperformed the baseline and default Cycle-GAN on Mean DSC on the STARE labelled data. 

\subsection{Results on CT/MR Domain Adaptation for MM-WHS LV and Myocardium Segmentation} 
The final segmentation results for the MR/CT MM-WHS Domain Adaptation approaches in this experiment are tabulated in Table \ref{tab:MRCTResults}. 

\begin{table}[h!]
    \centering
    \caption{Final segmentation results for Unsupervised MR/CT MM-WHS DA approaches. Best results highlighted in bold.}
    \begin{tabular}{c c c c c}

    \hline
    DA Direction & DA Method & Mean Overall DSC & LV DSC & Myo DSC\\
    
    \hline
    $CT \rightarrow MR$ & \textbf{SP Cycle-GAN} & \textbf{0.7820} $\pm$ 0.0051 & 0.8093 $\pm$ 0.0060 & \textbf{0.5574} $\pm$ 0.0102\\
    $CT \rightarrow MR$ & Default Cycle-GAN & 0.7612 $\pm$ 0.0051 & \textbf{0.8143} $\pm$ 0.0064 & 0.4939 $\pm$ 0.0094\\
    $CT \rightarrow MR$ & No Domain Adaptation & 0.7415 $\pm$ 0.0056 & 0.8102 $\pm$ 0.0078 & 0.4433 $\pm$ 0.0101\\
    \midrule
    $MR \rightarrow CT$ & \textbf{SP Cycle-GAN} & \textbf{0.8533} $\pm$ 0.0046 & \textbf{0.8310} $\pm$ 0.0106 & \textbf{0.7435} $\pm$ 0.0056\\
    $MR \rightarrow CT$ & Default Cycle-GAN & 0.8341 $\pm$ 0.0053 & 0.8145 $\pm$ 0.0108 & 0.7016 $\pm$ 0.0061\\
    $MR \rightarrow CT$ & No Domain Adaptation & 0.7908 $\pm$ 0.0055 & 0.7889 $\pm$ 0.0109 & 0.6047 $\pm$ 0.0071\\
    \hline
       
    \end{tabular}
    \label{tab:MRCTResults}
\end{table}

As can be seen in Table \ref{tab:MRCTResults}, for both the CR to MR and MR to CT directions, the SP Cycle-GAN outperforms the default Cycle-GAN and no DA approaches in terms of Mean DSC for nearly all classes. For the LV segmentation, the SP Cycle-GAN achieved better performance over other methods for MR to CT DA and marginally worse performance than the default Cycle-GAN for CT to MR (though the difference is very slight). For both DA directions however, a more significant segmentation performance increase can be seen in the Myocardium segmentation results for the SP Cycle-GAN over other methods. 

\section{Discussion}

\subsection{STARE and DRIVE Domain Adaptation}

Table \ref{tab:STAREDRIVEResults} summarizes the STARE and DRIVE DA results for binary blood vessel segmentation. For the STARE to DRIVE DA case, surprisingly, no DA approach outperformed the baseline method of simply training a segmentation model on the untranslated STARE data and testing on the DRIVE test dataset. This method achieved a mean DSC of 0.7706, marginally outperforming the SP Cycle-GAN method, which achieved a mean DSC of 0.7679. When considering the standard error of these measurements however, we can conclude that the SP Cycle-GAN had roughly the same performance as no domain adaptation for STARE to DRIVE. It is difficult to pinpoint why no DA approach could outperform the baseline approach for the STARE to DRIVE domain adaptation problem. One potential explanation is that the STARE domain blood vessels are intrinsically harder to segment than the DRIVE ones, so training a model on just the STARE data is sufficient for DRIVE testing. Additionally, the domain shift posed between STARE and DRIVE may not be sufficient for the Attention U-Net model to suffer greatly when not translating images first with a DA model. Also, the limited amount of DRIVE training data may make the learned target domain from the Cycle-GAN models not accurately representative of the DRIVE domain, leading to poorly translated images, ultimately hurting segmentation performance.  

\begin{figure}
  \centering
  \includegraphics[width=\linewidth]{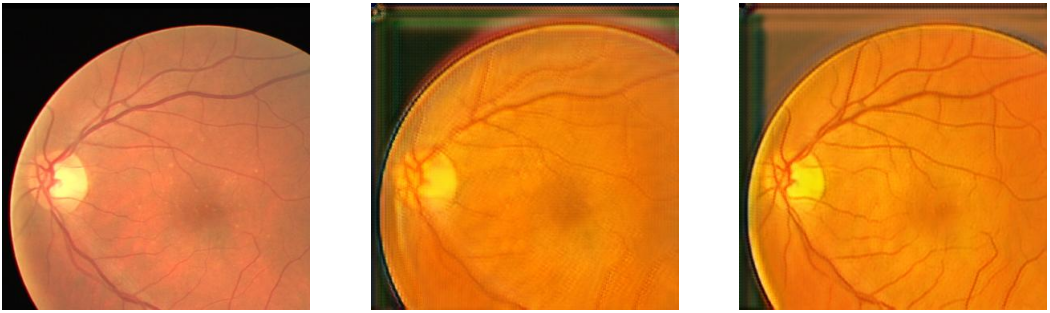}
  
  \caption{Demonstration of structure preserving ability of the SP Cycle-GAN for domain adaptation. Original DRIVE image patch (left) vs. default Cycle-GAN translated DRIVE to STARE patch (middle) vs. SP Cycle-GAN translated DRIVE to STARE patch (right).}
  \label{fig:SEGGAN_DEMO}
\end{figure}

Conversely, we see a significant segmentation performance increase for DRIVE to STARE DA for the SP Cycle-GAN, with a mean DSC of 0.7102 compared to 0.6714 and 0.6615 of the default Cycle-GAN and baseline methods. For DRIVE to STARE, both DA approaches outperform the baseline, indicating that the domain shift posed by exposure to the STARE domain after DRIVE training is significant enough to slightly hurt performance. The SP Cycle-GAN clearly improves performance over the default Cycle-GAN, showing the SP Cycle-GANs effectiveness in both addressing the overall domain shift and maintaining the blood vessel structure after image translation. This difference in performance indicates that while the default Cycle-GAN can effectively translate the DRIVE images to the STARE domain, it is not doing a good job of preserving the overall structure of the blood vessels which are later segmented. The structure preserving ability of the SP Cycle-GAN compared to the default Cycle-GAN is highlighted in Figure \ref{fig:SEGGAN_DEMO}. 

As can be seen in this figure, the default Cycle-GAN translated image blurs the blood vessel structure and features blood vessel structures which are difficult to see when compared to the original DRIVE image. Conversely, the SP Cycle-GAN translated patch maintains the blood vessel structure exceptionally well, and even seems to increase the contrast between the blood vessel structure and the background of the scan. This visual evidence of structure preservation is backed up by the improved performance of the SP Cycle-GAN over other methods for DRIVE to STARE domain adaptation. 

\subsection{MM-WHS MR and CT Domain Adaptation}
Table \ref{tab:MRCTResults} shows the MR and CT DA results for the MM-WHS dataset. The SP Cycle-GAN shows a clear performance increase for both CT to MR and MR to CT DA in terms of mean multi-class DSC over all other methods. As can be noted for both directions, the largest segmentation improvement the SP Cycle-GAN provides is the difficult to segment Myocardium class. For CT to MR, the SP Cycle-GAN has a Myocardium DSC of 0.5574, while the default Cycle-GAN and baseline methods can only achieve 0.4939 and 0.4433 respectively. This trend continues for the MR to CT direction, with the SP Cycle-GAN achieving a mean Myocardium class DSC of 0.7435, compared to a DSC of 0.7016 and 0.6047 achieved by the default Cycle-GAN and baseline methods respectively. The Myocardium is a difficult to segment region of the heart for the MM-WHS dataset, with Dice scores for this class being signficantly lower on average than LV segmentation in other experiments \cite{CFDistance}. The SP Cycle-GAN achieves state of the art segmentation performance on the Myocardium for the MR to CT direction compared to Wu and Zhuang (2020), who achieved a Myocardium Dice score of 0.679 for their method. For the MR to CT direction, the LV segmentation also outperforms other methods, with a mean DSC of 0.8310. The only metric in which the SP Cycle-GAN does not outperform the other methods for the MM-WHS dataset is the LV DSC for the CT to MR direction, achieving an LV DSC of 0.8093 compared to the 0.8143 of the default Cycle-GAN. However, when factoring the standard errors for these measurements, the performance can be deemed fairly even for LV segmentation. 

The SP Cycle-GAN shows a clear ability to preserve the structure of the difficult to segment Myocardium region in the MM-WHS DA problems, and achieves excellent performance in Myocardium and overall multi-class segmentation compared to other methods, and the state of the art 2D MM-WHS segmentation scores \cite{CFDistance}. 

\subsection{Machine Learning and Healthcare Implications}

As demonstrated above, the introduction of a simple segmentation loss term into the Cycle-GAN training process, seen in Equation \ref{eq:SEGGANLoss}, which forms the SP Cycle-GAN, allows for a domain adaptation method for medical images which better preserves the structure of medical regions of interest. This is backed up by visual evidence from the translated scans, seen in Figure \ref{fig:SEGGAN_DEMO}, and the significant performance increase achieved by the SP Cycle-GAN for the DRIVE to STARE, MM-WHS CT to MR and MM-WHS MR to CT domain adaptation problems. This segmentation loss is extendable to other machine learning implementations of medical image domain adaptation for segmentation, and can potentially bolster a domain adaptation models ultimate segmentation performance with the simple inclusion of a segmentation loss term in the DA model loss. Adversarial domain adaptation models can benefit from the inclusion of this structure-preserving loss term, and create unsupervised domain adaptation models which better aid clinicians in multi-modal segmentation for diagnoses from medical scans.

\paragraph{Limitations}
The main limitation of the approach of the SP Cycle-GAN method for structure-preserving adversarial domain adaptation, is the requirement of labelled source domain images. Without labelled source domain images, featuring a binary or multi-class segmentation map of regions of interest in the image, the loss formulated in Equation \ref{eq:SEGGANLoss}, cannot be implemented. This is not the case of other similar structure-preserving unsupervised DA techniques like the CannyEdgeGAN \cite{CANNY-GAN} or CFD \cite{CFDistance}. Another limitation of this methodology is the limited data available for the STARE and DRIVE datasets for binary blood vessel segmentation \cite{STARE,DRIVE}. Unlike the DRIVE dataset which features a seperate training and testing set of labelled data, the STARE dataset features only 20 labelled images out of a total of 397 scans \cite{STARE}. The requirement to use labelled STARE images for the SP Cycle-GAN training may limit its capacity to learn the STARE image domain without being able to access the other over 300 unlabelled images during training. 

\section{Conclusion}
In this work, we introduced SP Cycle-GAN, which uses simultaneous source domain segmentation during Cycle-GAN training to enforce a structure preserving loss term in the Cycle-GAN DA model. The inclusion of this structure preserving loss term resulted in a Cycle-GAN based DA model which effectively preserved blood vessel structure during STARE and DRIVE retinal scan domain adaptation, both visually and in terms of final Dice segmentation scores, when compared to a standard Cycle-GAN DA approach. Additionally, the SP Cycle-GAN model outperformed baseline and Cycle-GAN implementations for MR and CT domain adaptation in the MM-WHS whole heart segmentation dataset for Left Ventricle and Myocardium segmentation. The SP Cycle-GAN showed a very strong capacity to preserve the translated Myocardium structure (which is a difficult to segment region when compared to the LV) compared to other methods, and achieved state of the art Myocardium segmentation results. The developed SP Cycle-GAN methodology shows promise to improve the accuracy and reliability of unsupervised adversarial DA models for medical image segmentation purposes. 

\section*{Acknowledgments}
This work was supported in part by the TMU Multimedia Research Laboratory of Toronto Metropolitan University. 

\bibliographystyle{unsrt}  
\bibliography{references}

\end{document}